# Three State Quantum System Exhibiting Third Order Exceptional Singularities and Flip-of-States

Sayan Bhattacherjee, Arnab Laha, and Somnath Ghosh*

[1]Department of Physics, Indian Institute of Technology Jodhpur, Rajasthan-342037, India
*somiit@rediffmail.com

**Abstract:** A quantum inspired open optical system is mathematically implemented with an analogous three state non-Hermitian Hamiltonian exhibiting two special avoided-resonance-crossings; where interesting characteristics alongside a third-order exceptional point is explored towards robust ultra-selective state switching.

## 1. Introduction

Lately, many enticing ideas of non-Hermitian quantum mechanics have inspired the various integrated optical phenomena, where appearance of exceptional singularities, namely exceptional points ($EP$) and its fascinating physical effects have been attracted immense attention; and also discussed extensively [1, 2] in various open optical systems like coupled waveguides [3], gain-loss microcavities [4], etc. A second order $EP$ ($EP2$) is the particular point in $2D$ parameter spectrum (at least two real or a complex parameter) where two eigenvalues and also the corresponding eigenstates coalesce and connected by branch point singularities. The location of a $EP2$ is referred by a special avoided resonance crossing ($ARC$) between two coupled eigenvalues with crossing/ anti-crossing of their real/ imaginary parts. In this paper, we propose a three state non-Hermitian Hamiltonian where a specific eigenvalue is coupled with both of the rest eigenvalues individually which correspond two interacting $EP2$s. Investigating such interactions between two $EP2$s in chosen parameter plane, we realize the existence of a third-order $EP$ ($EP3$). Recently, $EP3$ have been widely studied and implemented as convenient tool to explore wide varieties of intriguing phenomena like resonance scattering [5], extreme enhanced sensing [6], etc. Here, we exclusively report successive sate-switching among three interacting states around an $EP3$ for the first time. Such successive state exchange is an important consequence of existence of an $EP3$ [3].

## 2. Non-Hermitian Mathematical Model and Numerical Results

To study the phenomena of three-state-interaction via two special $ARC$s, a simplest form of a $3 \times 3$ non-Hermitian Hamiltonian $\mathcal{H}$ having form $H_0 + \lambda H_p$ is presented as follows.

$$\mathcal{H} = \begin{pmatrix} \widetilde{\varepsilon}_1 & 0 & 0 \\ 0 & \widetilde{\varepsilon}_2 & 0 \\ 0 & 0 & \widetilde{\varepsilon}_3 \end{pmatrix} + \lambda \begin{pmatrix} 0 & \omega_{12} & 0 \\ \omega_{21} & 0 & \omega_{23} \\ 0 & \omega_{32} & 0 \end{pmatrix}; \text{ where } \widetilde{\varepsilon}_j = \varepsilon_j + \delta_j, (j = 1, 2, 3) \quad (1)$$

In the passive portion $H_0$, $\varepsilon_j$ ($j = 1,2,3$) are the real passive eigenvalues with small complex displacements $\delta_j$ (as $\delta_j = \delta_{jR} + i\delta_{jI}$, $\delta_{jI}$ essentially comparable with decay rates); which are being interacted by the coupling matrix $H_p$. Here the elements of $H_p$ are judicially customized as $\omega_{12} = (i + \widetilde{\varepsilon}_1 - \widetilde{\varepsilon}_2 - \widetilde{\varepsilon}_3)$, $\omega_{21} = (\widetilde{\varepsilon}_2 - \widetilde{\varepsilon}_3)$, $\omega_{23} = (\widetilde{\varepsilon}_2 + \widetilde{\varepsilon}_3)$ and $\omega_{32} = 0.5i + \Re(\widetilde{\varepsilon}_1 - \widetilde{\varepsilon}_3)$. During optimization, we choose $\varepsilon_1 = 0.7$, $\varepsilon_2 = 0.65$ and $\varepsilon_3 = 0.3$. $\delta_{jI}$ ($j = 1,2,3$) are fixed at $0.25$. We also fix $\delta_{2R} = 0 = \delta_{3R}$. Now the coupling phenomena between the eigenvalues $E_j$ ($j = 1,2,3$) of $\mathcal{H}$ are controlled by tuning $\delta_{1R}$ over an independent complex parameter $\lambda$ (as $\lambda = \lambda_R + i\lambda_i$).

In Fig. 1(a), the interactions between $E_1$ and $E_3$ (unaffecting $E_2$) are demonstrated via a special $ARC$ in complex eigenvalue plane ($E$-plane) with variation in $\lambda$ (where $\lambda_R$ varies from 0 to -1 with simultaneous tunable $\lambda_I$ from 0 to -0.6). For $\delta_{1R} = 0.003$, they exhibit $ARC$ (depicted in Fig. 1(a.1); directed by brown dotted arrows) with crossing in $\Re[E]$ and anti-crossing in $\Im[E]$ w.r.t. $\lambda_R$ (shown in Fig. 1(a.2)). Now for slightly higher value of $\delta_{1R} = 0.004$, a different kind of $ARC$ occurs (shown by black solid arrows in Fig. 1(a.1)); where $\Re[E]$ experiences anti-crossing with simultaneous crossing in $\Im[E]$ w.r.t. $\lambda_R$ as shown in Fig. 1(a.3). This abrupt behavioral change in $ARC$ for two different $\delta_{1R}$ clearly indicates the presence of an $EP2$ in $(\delta_{1R}, \delta_{1I})$-plane at $\sim (0.0035, 0.25)$ [3]. Similarly, such special $ARC$ between $E_1$ and $E_2$ (unaffecting $E_3$) is exhibited in Fig. 1(b.1); where dissimilar behavior in crossing/ anti-crossing of corresponding $\Re[E]$ and $\Im[E]$ for chosen $\delta_{1R}$ as 0.12 and 0.125 as shown in Fig. 1(b.2) and (b.3) respectively indicates the presence of another $EP2$ at $(\delta_{1R} = 0.1225, \delta_{1I} = 0.25)$. Thus, we behold two situations w.r.t. choice of $\delta_{1R}$ over the specified $\lambda$-span, where among three interacting states any two are interacting keeping the rest as observer; while interacting states exhibit special $ARC$ and coalesce at an $EP2$. The whole phenomena are presented in Fig. 1(c). This is an exclusive signature of the presence of an EP3 [2].

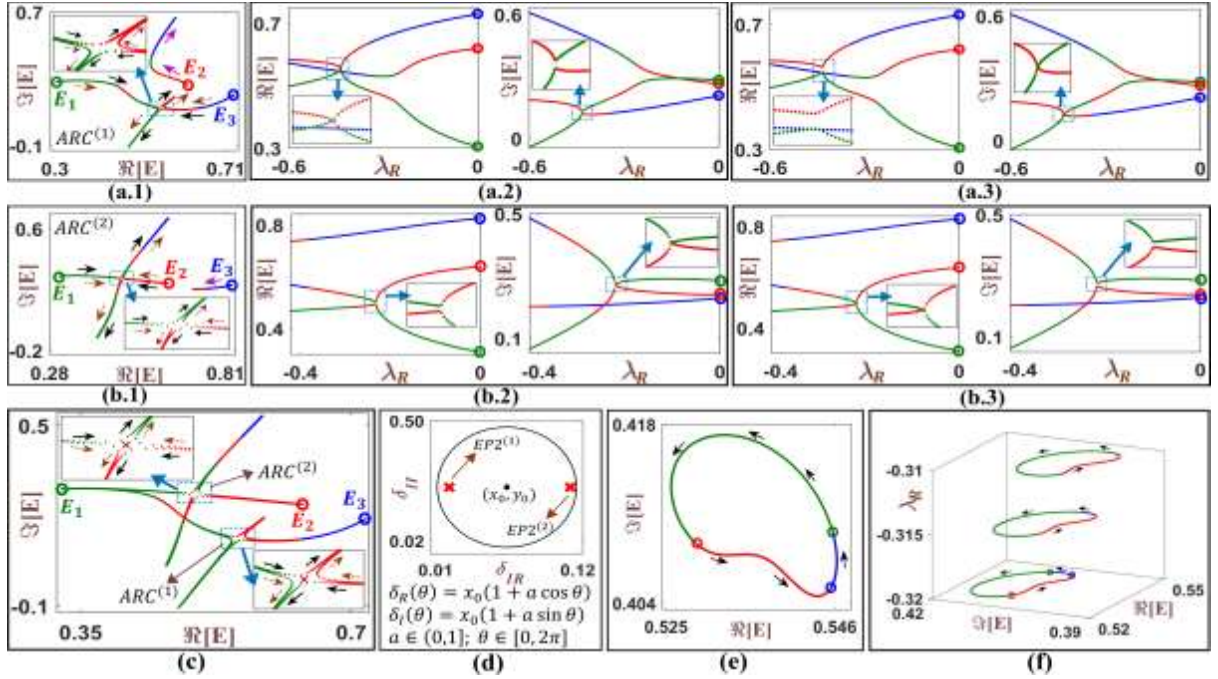

**Fig. 1:** **(a)** Trajectories of $E_j$ (shown by green, red and blue dots for $j=1,2,3$ respectively) with $\lambda$ when **(a.1)** $E_1$ and $E_3$ exhibit $ARCs$ (unaffecting $E_2$) with corresponding **(a.2)** crossing and anti-crossing in $\Re[E]$ and $\Im[E]$ w.r.t. $\lambda_R$ for $\delta_{1R} = 0.003$; and **(a.3)** vice-versa for $\delta_{1R} = 0.004$ respectively. **(b)** Similar trajectories when **(b.1)** $E_1$ and $E_2$ exhibit $ARCs$ (unaffecting $E_3$) with corresponding **(b.2)** anti-crossing and crossing in $\Re[E]$ and $\Im[E]$ for $\delta_{1R} = 0.12$; and **(b.3)** vice-versa for $\delta_{1R} = 0.125$ respectively. The regions marked by dotted rectangles are zoomed in respective insets. **(c)** Three-state-$ARC$ associated with all the three interacting states at a time; where in the zoomed portions red crosses indicate the approximate equivalent positions of two $EP2s$ in complex $E$-plane. **(d)** Enclosing two identified $EP2s$ in $(\delta_{1R}, \delta_{1I})$-plane described by the given set of parametric Eqs. of a circle with center at $(x_0 = 0.063, y_0 = 0.25)$ and radius $a\,(=0.98)$. $EP2^{(1)}$ and $EP2^{(2)}$ are corresponds to $ARC^{(1)}$ and $ARC^{(2)}$ as shown in (a.1) and (b.1) respectively. **(e)** Complex trajectories of all three interacting eigenvalues exhibiting successive state switching followed by an anti-clockwise variation of $\delta_{1R}$ and $\delta_{1I}$ along the closed contour described in (d) for a fixed $\lambda_R$. **(f)** Robustness of flip-of-states phenomena as described in (e) w.r.t. $\lambda_R$. Arrows indicates the direction of progressions.

Now we study the unconventional physical effects near/ around $EP3$ via considering the effect of encircling around two identified $EP2s$ in $(\delta_{1R}, \delta_{1I})$-plane as presented in Fig. 1(d). Now following the progressive circular variation of $\delta_{1R}$ and $\delta_{1I}$ along the described closed loop, three interacting states are exchanging their positions successively. Such successive switching, namely flip-of-states for a fixed $\lambda$ is displayed in Fig. 1(e). Now in Fig. 1(f), we present the robustness of the described flip-of-states w.r.t. $\lambda_R$; here as can be seen, the successive flip-of-states is omnipresent even in variation of $\lambda_R$.

## 4. Summary

In summary, a three state non-Hermitian Hamiltonian matrix, hosting three interacting eigenvalues is proposed and modelled. From the optimization and parametric dependence, one may analogically realize the interactions between the quantum states associated with an open optical system. The proposed Hamiltonian exhibits two special ARCs associated with three interacting states in complex eigenvalue-plane which corresponds two EP2s in system parameter plane. We study the interaction between two identified EP2s via scanning the alongside areas at a time by enclosing them properly inside a closed contour and realize the presence of an EP3 in the chosen parameter plane. With such encirclement, the successive switching among three corresponding states in complex eigenvalue-plane is reported for the first time towards robust flip-of-states phenomenon.

AL and SG acknowledges support from DST, India [IFA-12; PH-23]. SB acknowledges support from MHRD.